\documentclass[aps,12pt,final,notitlepage,oneside,onecolumn,%
nobibnotes,nofootinbib,superscriptaddress,noshowpacs,centertags]%
{revtex4}

\usepackage{graphicx}

\begin{document}

\title{{ Once more on electromagnetic form factors of nucleons in extended
vector meson dominance model} }

\author{\firstname{Amand} \surname{Faessler}}
\affiliation{Institut f\"ur Theoretische Physik der Universit\"at T\"ubingen, 
Auf der Morgenstelle 14, D-72076 T\"ubingen, Germany}
\author{\firstname{M.I.} \surname{Krivoruchenko}}
\affiliation{Institut f\"ur Theoretische Physik der Universit\"at T\"ubingen, 
Auf der Morgenstelle 14, D-72076 T\"ubingen, Germany}
\affiliation{Institute for Theoretical and Experimental
Physics, B. Cheremushkinskaya 25, 117259 Moscow, Russia}
\author{\firstname{B.V.} \surname{Martemyanov}}
\affiliation{Institut f\"ur Theoretische Physik der Universit\"at T\"ubingen, 
Auf der Morgenstelle 14, D-72076 T\"ubingen, Germany}
\affiliation{Institute for Theoretical and Experimental
Physics, B. Cheremushkinskaya 25, 117259 Moscow, Russia}

\begin{abstract}
Extended vector meson dominance model, that allows to describe the
electromagnetic form factors of nucleons obeying the asymptotic quark
counting rule prescriptions and contains the minimal number of free
parameters, is presented. We get a reasonable fit of form factors over
experimentally available space-like region of momentum transfer and get also
reasonable results in the time-like region.
\end{abstract}

\maketitle


\section{Introduction}

In a series of our papers an extended vector meson dominance model was successfully 
applied to the description of  electromagnetic transition
form factors of nucleon resonances~ \cite{eVMD} that are necessary to find out the dilepton outcome from the decays
of these resonances produced either in proton-proton~ \cite{ppDLS} or in heavy ion collisions
~\cite{HICDLS, HICHADES}. We
didn't say anithing about how the same model could describe the fundamental elecrtomagnetic
form factors of nucleons themselves. Although these form factors were the subject of
considerations of many other models (see the review~\cite{perd} for the elongeted list of references) 
including the extended vector meson dominance model itself
\cite{duby} 
we would like to apply our own version of extended vector meson dominance model to
the description of nucleon form factors. It contains the minimal number of free
parameters and gives a reasonable result.

\section{Nucleon electromagnetic form factors in the space-like region}

Electromagnetic form factors of nucleon $F_1^N(q^2)$, $F_2^N(q^2)$ describe
 its electromagnetic current 
\begin{equation} <N(p^\prime)|J_\mu|N(p)> = {\bar u}(p^\prime)(F_1^N(q^2)\gamma_\mu +\sigma_{\mu\nu}\frac{q_\nu}{2m}F_2^N(q^2))u(p)\end{equation}
and define the elastic scattering of electrons off nucleons in the space-like region $q^2=-Q^2<0$
and electron-positron annihilation to nucleon-antinucleon in the time-like region $q^2>0$.
The values  of form factors at origin $F_1^N(0)$, $F_2^N(0)$ are connected to the charges and anomalos magnetic moments of
nucleons and are equal for proton and neutron to
\begin{equation} F_1^p(0) = 1,~~~F_2^p(0) = 1.79,~~~F_1^n(0) = 0,~~~F_2^n(0) = -1.92~.\end{equation}

Quark counting rules ~\cite{VaZa} predict the asymptotics of form factors in the space-like region $Q^2\rightarrow\infty$
\begin{equation}F_1^N(Q^2) \sim \frac{1}{Q^4},~~~F_2^N(Q^2) \sim \frac{1}{Q^6}~.\end{equation}
In extended Vector Meson Dominance (eVMD) model such asymptotics comes from destructive interference
of ground and excited states of vector mesons $V,~V^\prime,~V^{\prime\prime}$
\begin{equation}F_1^N(Q^2) = \frac{F_1^N(0)+c^N Q^2}{(1+\frac{Q^2}{m^2_V})(1+\frac{Q^2}{m^2_{V^\prime}})(1+\frac{Q^2}{m^2_{V^{\prime\prime}}})},
~F_2^N(Q^2) = \frac{F_2^N(0)}{(1+\frac{Q^2}{m^2_V})(1+\frac{Q^2}{m^2_V{^\prime}})(1+\frac{Q^2}{m^2_{V^{\prime\prime}}})}~.\end{equation}

Isovector $F_1^\rho(Q^2)$, $F_2^\rho(Q^2)$ and isoscalar $F_1^\omega(Q^2)$, $F_2^\omega(Q^2)$ form factors
are defined as linear superpositions of proton and neutron form factors
\begin{equation}F_{1,2}^\rho(Q^2) = \frac{F_{1,2}^p(Q^2)-F_{1,2}^n(Q^2)}{2},~~~F_{1,2}^\omega(Q^2) = \frac{F_{1,2}^p(Q^2)+F_{1,2}^n(Q^2)}{2}~\end{equation}
and
are described the  coupling constants of $\rho$ and $\omega$ family mesons to nucleon ($ f_{1,2}^{\rho NN}$, ...)and to photon 
($ g_\rho$, ...)respectively
\begin{equation}F_{1,2}^\rho(Q^2) = \frac{f_{1,2}^{\rho NN}}{g_\rho}\frac{m_\rho^2}{m_\rho^2+Q^2}+\frac{f_{1,2}^{{\rho^\prime} NN}}{g_{\rho^\prime}}\frac{m_{\rho^\prime}^2}{m_{\rho^\prime}^2+Q^2}+\frac{f_{1,2}^{{\rho^{\prime\prime}} NN}}{g_{\rho^{\prime\prime}}}\frac{m_{\rho^{\prime\prime}}^2}{m_{\rho^{\prime\prime}}^2+Q^2}\end{equation}
\begin{equation}F_{1,2}^\omega(Q^2) = \frac{f_{1,2}^{\omega NN}}{g_\omega}\frac{m_\omega^2}{m_\omega^2+Q^2}+\frac{f_{1,2}^{{\omega^\prime} NN}}{g_{\omega^\prime}}\frac{m_{\omega^\prime}^2}{m_{\omega^\prime}^2+Q^2}+\frac{f_{1,2}^{{\omega^{\prime\prime}} NN}}{g_{\omega^{\prime\prime}}}\frac{m_{\omega^{\prime\prime}}^2}{m_{\omega^{\prime\prime}}^2+Q^2}.\end{equation}

 The masses of $\rho$ and $\omega$ family mesons are assumed to be the same and this allows to set an equivalence
between the multiplicative forms (4) and additive forms (following from eqs. (5) -- (7)) for proton and neutron
form factors.

With the choice $m_\rho=m_\omega=0.770$ GeV, $m_{\rho^\prime}=m_{\omega^\prime}=1.250$ GeV, $m_{\rho^{\prime\prime}}=m_{\omega^{\prime\prime}}=1.450$ GeV (as was used before in the case
 of electromagnetic transition
form factors of nucleon resonances~ \cite{eVMD})  we have two free parameters to describe
form factors in the space-like region: $c^p$ and $c^n$. They were fitted and are equal to  $c^p=0.463$ GeV$^{-2}$
and $c^n=-0.297$ GeV$^{-2}$. The results of the fit are presented in $q^2=-Q^2<0$ regions of the Fig.1 and Fig.2
where electric and magnetic form factors
\begin{equation}G_E^N = F_1^N +\frac{q^2}{4m^2}F_2^N, ~~~~~ G_M^N = F_1^N +F_2^N\end{equation} are shown.
At small $Q^2$ the decomposition $G_E^N \approx F_1^N(0) - \frac16 Q^2<r_N^2>$ defines the charge radii of
proton $\sqrt{<r_p^2>} = 0.83$ fm (exp: 0.875 fm) and neutron $<r_n^2> = -0.06$ fm$^2$ (exp: -0.113 fm$^2$).

For known coupling constants of the photon to $\rho$ and $\omega$ mesons $g_\rho = 5.03$ and $g_\omega = 17.1$
their coupling constants to the nucleon are equal to
\begin{equation}f_1^{\rho NN} = 3.02,~~~f_2^{\rho NN} = 20.8,~~~f_1^{\omega NN} = 17.2,~~~f_2^{\omega NN} = -2.47\end{equation}
what is close to corresponding  coupling constants used to describe Bonn potential \cite{Mach} of nucleon-nucleon
interaction: $f_1^{\rho NN} = 3.2,~~~f_2^{\rho NN} = 19.8,~~~f_1^{\omega NN} = 15.9,~~~f_2^{\omega NN} = 0$.

\section{Nucleon electromagnetic form factors in the time-like region}

In time-like region the finite widths of vector mesons should be taken into account
$$F_{1,2}^\rho(q^2) = \frac{f_{1,2}^{\rho NN}}{g_\rho}\frac{m_\rho^2}{m_\rho^2-im_\rho\Gamma_\rho-q^2}+\frac{f_{1,2}^{{\rho^\prime} NN}}{g_{\rho^\prime}}\frac{m_{\rho^\prime}^2}{m_{\rho^\prime}^2-im_{\rho^\prime}\Gamma_{\rho^\prime}-q^2}$$$$+\frac{f_{1,2}^{{\rho^{\prime\prime}} NN}}{g_{\rho^{\prime\prime}}}\frac{m_{\rho^{\prime\prime}}^2}{m_{\rho^{\prime\prime}}^2-im_{\rho^{\prime\prime}}\Gamma_{\rho^{\prime\prime}}-q^2}$$
$$F_{1,2}^\omega(q^2) = \frac{f_{1,2}^{\omega NN}}{g_\omega}\frac{m_\omega^2}{m_\omega^2-im_\omega\Gamma_\omega-q^2}+\frac{f_{1,2}^{{\omega^\prime} NN}}{g_{\omega^\prime}}\frac{m_{\omega^\prime}^2}{m_{\omega^\prime}^2-im_{\omega^\prime}\Gamma_{\omega^\prime}-q^2}$$$$+\frac{f_{1,2}^{{\omega^{\prime\prime}} NN}}{g_{\omega^{\prime\prime}}}\frac{m_{\omega^{\prime\prime}}^2}{m_{\omega^{\prime\prime}}^2-im_{\omega^{\prime\prime}}\Gamma_{\omega^{\prime\prime}}-q^2}$$

We took $\Gamma_\rho=0.150$ GeV, $\Gamma_\omega=0.0085$ GeV, $\Gamma_{\rho^\prime} = \Gamma_{\omega^\prime} =0.300$ GeV,
$\Gamma_{\rho^{\prime\prime}} = \Gamma_{\omega^{\prime\prime}} = 0.500$ GeV. The exact values of the widths are not very important
when we are far from the resonance region. Figs.1,2 show the prediction of eVMD model for time-like $q^2>0$.
Experimental data were obtained with the assumption $|G_E^p| = |G_M^p|$ in the case of proton and with the assumption $|G_E^n| = 0$
in the case of neutron\cite{voci}. Theoretically we have $|G_E^N| > |G_M^N|$ and their ratio grows
with $q^2$. This means that for large $q^2$ the experimental points for $|G_M^N|$ should go significantly lower
what is appreciated. Our results on $|G_E^p|/|G_M^p|$ ratio can be considered as the predictions for the future (like FAIR) antiproton facilities.

In the unphysical region of $q^2$, $0 < q^2 < 4m^2$, indicated on Figs.1,2 by vertical lines the resonance peaks of vector mesons (the largest one of $\omega$ meson)
are clearly seen.

\section{Conclusion} An extended vector meson dominance model with a minimal number of free parameters
 is applied to the description of electromagnetic  form factors of nucleons. The couplings
of ground state $\rho$ and $\omega$ mesons to the nucleons are calculated and appear to be close to those
of Bonn potential model of nucleon interaction. In the time-like region the absolute values of electric form factors are
considerably larger than those of magnetic form factors and this can be used in the reanalysis of experimental
data obtained with the assumption $|G_E^p| = |G_M^p|$  in the proton case and  $|G_E^n| = 0$ in the neutron case.

\acknowledgments
This work is supported by RFBR grant No.
09-02-91341 and DFG grant No. 436 RUS 113/721/0-3. 
M.I.K. and  B.V.M.  acknowledge the kind hospitality at the University
 of T\"{u}bingen.

\newpage

\begin{figure}[!htb]
\begin{center}

\vspace{-2 cm}
\includegraphics[width=0.8\textwidth]{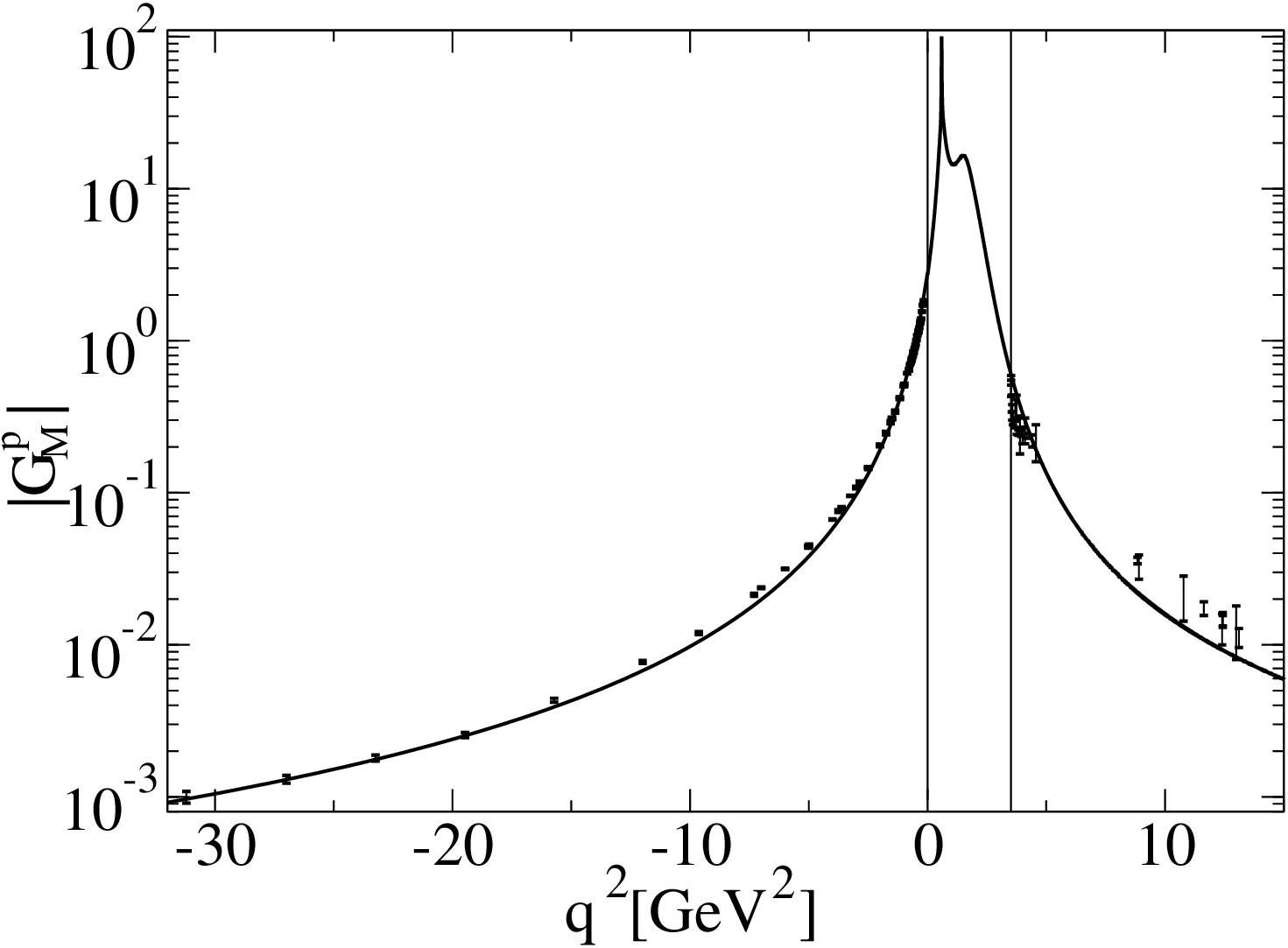}

\vspace{1 cm}
\includegraphics[width=0.8\textwidth]{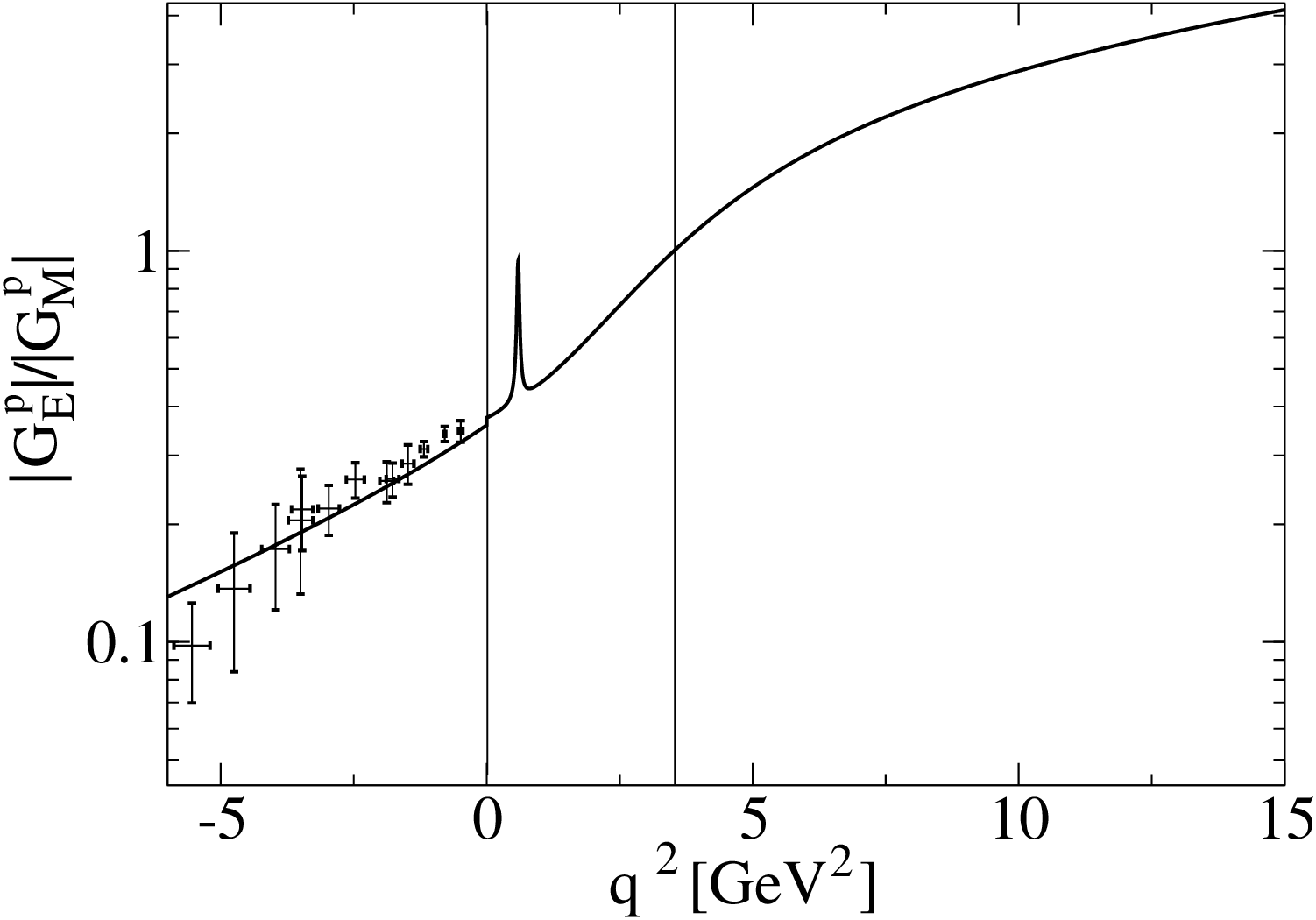}
\end{center}
\caption {The modulus of magnetic form factor (up) and of the ratio of electric to magnetic form factors (down) 
of the proton in space- and time-like regions (for space-like region we use experimental data cited in review \cite{perd}).}
\end{figure}
\begin{figure}[!htb]
\begin{center}

\vspace{-2 cm}
\includegraphics[width=0.8\textwidth]{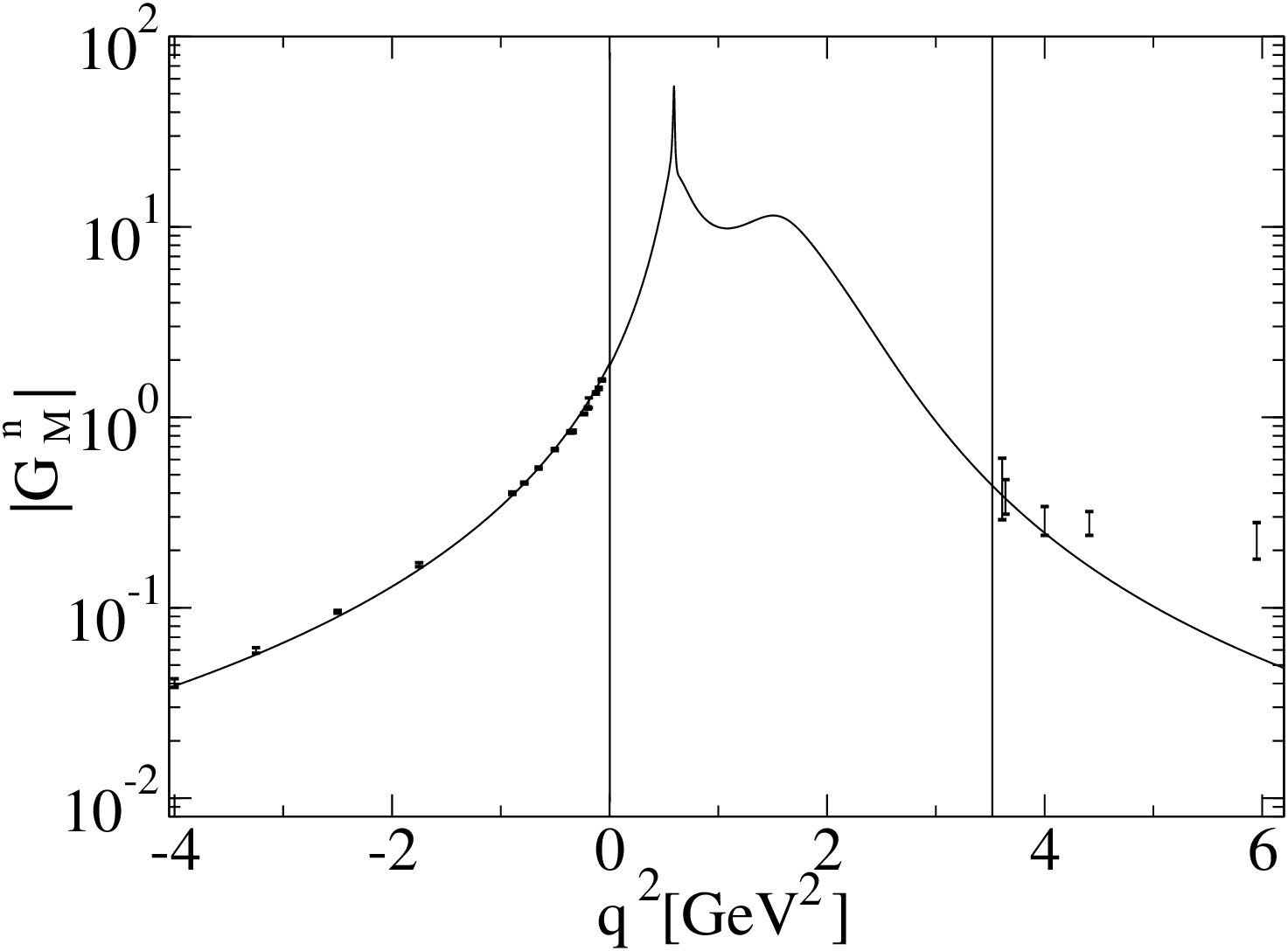}

\vspace{1 cm}
\includegraphics[width=0.8\textwidth]{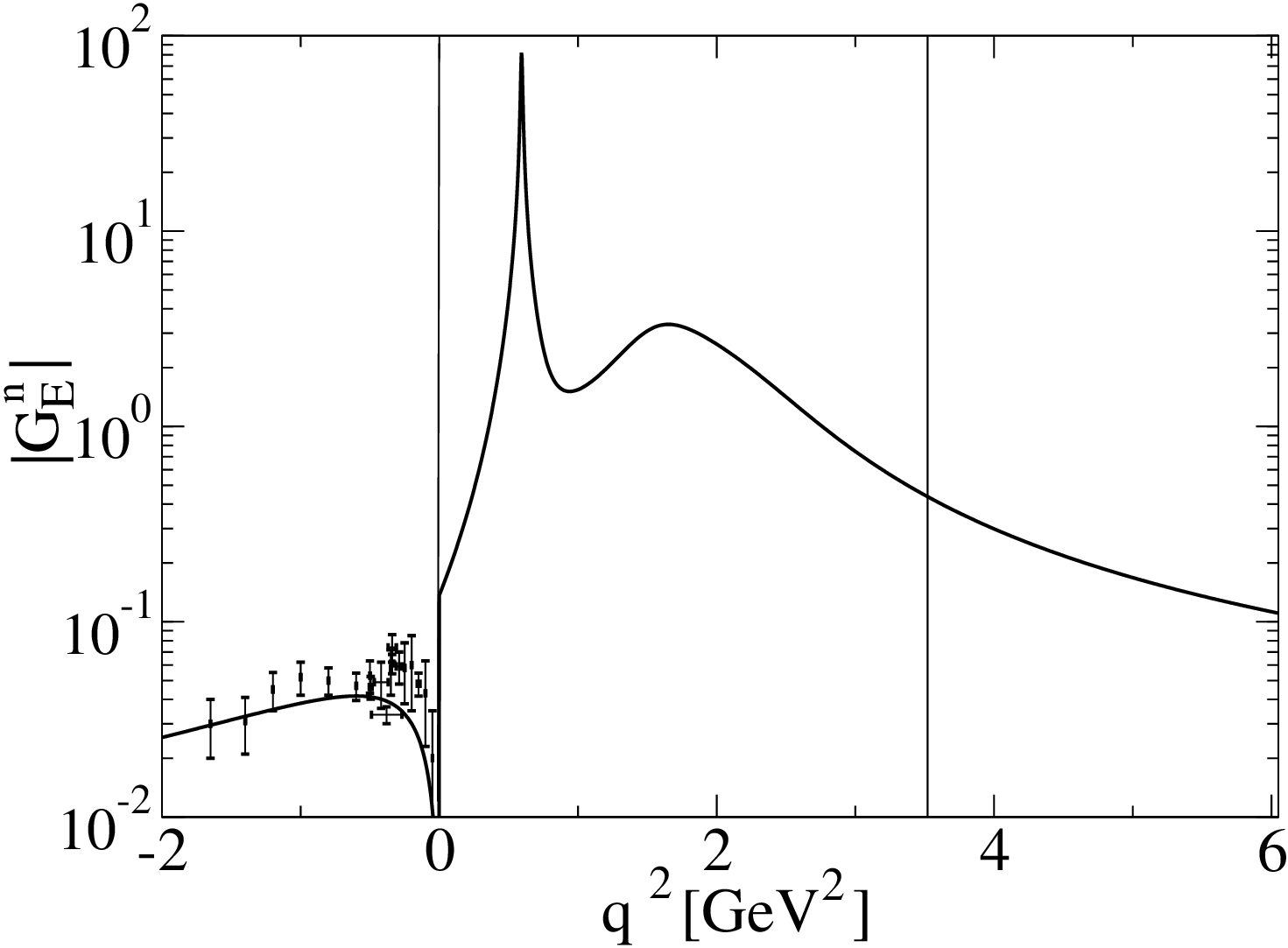}
\end{center}
\caption {The modulus of magnetic form factor (up) and the modulus of electric form factor (down) 
of the neutron in space- and time-like regions (for space-like region we use experimental data cited in review \cite{perd}).}
\end{figure}

\end{document}